\input  phyzzx
\input epsf
\overfullrule=0pt
\hsize=6.5truein
\vsize=9.0truein
\voffset=-0.1truein
\hoffset=-0.1truein

%
%

\def\IC{{\ \hbox{{\rm I}\kern-.6em\hbox{\bf C}}}}
\def\IR{{\hbox{{\rm I}\kern-.2em\hbox{\rm R}}}}
\def\IZ{{\hbox{{\rm Z}\kern-.4em\hbox{\rm Z}}}}

\def\sIR{{\hbox{{\sevenrm I}\kern-.2em\hbox{\sevenrm R}}}}

%
%
\hyphenation{Min-kow-ski}

\rightline{SU-ITP-97-25}
\rightline{May  1997}
\rightline{hep-th/9705107}

\vfill

%
%
\title{\bf M(atrix) Black Holes in Five Dimensions}

\bigskip

%
%

\author{Edi Halyo\foot{e-mail: halyo@dormouse.stanford.edu}}

\medskip

\address{Department of Physics \break Stanford University \break
 Stanford, CA 94305-4060}

\vfill

%
%

We examine five dimensional extreme black holes with three charges in the
matrix model. We build configurations of the $5+1$ super Yang--Mills theory
which correspond to black holes with transverse momentum charge. We calculate
their mass and entropy from the super Yang--Mills theory and find that they
match the semi--classical black hole results. We extend our results to
nonextreme black holes in the dilute gas approximation.

\vfill\endpage

%
%

\REF\SV{A. Strominger and C. Vafa, Phys. Lett. {\bf B379} (1996) 99,
hep-th/9601029; A. Tseytlin,  Mod. Phys. Lett. {\bf A11} (1996) 689,
hep-th/9601177.}
\REF\CM{ C. Callan and J. Maldacena, Nucl. Phys. {\bf B472} (1996) 591,
hep-th/9602043;
M. Cvetic and D. Youm, hep-th/9603100.}
\REF\DB{J. Polchinski, Phys. Rev. Lett. {\bf 75} (1997) 4724, hep-th/9510017;
hep-th/9611050;
J. Polchinski, S. Chaudhuri and C. Johnson, hep-th/9602052 and references
therein.}
\REF\HMS{G. Horowitz, J. Maldacena and A. Strominger, Phys. Lett. {\bf B383}
(1996) 151, hep-th/9603109.}
\REF\MS{J. Maldacena and L. Susskind, Nucl. Phys {\bf B475} (1996) 679,
hep-th/9604042.}
\REF\MAL{J. Maldacena, hep-th/9611125.}
\REF\DM{A. Dhar, G. Mandal and S. Wadia, hep-th/9605234; S. Das and S. Mathur,
Nucl. Phys. {\bf B478} (1996) 561, hep-th/9606185; hep-th/9607149.}
\REF\SCA{B. Kol and A. Rajaraman, hep-th/9608126; C. Callan, S. Gubser, I.
Klebanov and A. Tseytlin, hep-th/9610172; S. Gubser and I. Klebanov, Phys. Rev.
Lett. {\bf77} (1996) 4491, hep-th/9609076.}
\REF\KT{I. Klebanov and A. Tseytlin, Nucl. Phys. {\bf B475} (1996) 179,
hep-th/9604166.}
\REF\BFSS{T. Banks, W. Fischler, S. Shenker and L. Susskind, hep-th/9610043.}
\REF\WIT{E. Witten, Nucl. Phys. {\bf B460} (1996) 335, hep-th/9510135.}
\REF\WATI{W. Taylor, hep-th/9611042.}
\REF\LEN{L. Susskind, hep-th/9704080}
\REF\HS{G. Horowitz and A. Strominger, Phys. Rev. Lett. {\bf 77} (1996) 2368,
hep-th/9602051.}
\REF\DVV{R. Dijkgraaf, E. Verlinde and H. Verlinde, hep-th/9704018.}
\REF\LM{M. Li and E. Martinec, hep-th/9703211; hep-th/9704134.}
\REF\BRS{M. Berkooz, M. Rozali and N. Seiberg, hep-th/9704089.}
\REF\FHRS{W. Fischler, E. Halyo, A. Rajaraman and L. Susskind, hep-th/9703102.}
\REF\EDI{E. Halyo, hep-th/9704086.}
\REF\ST{L. Susskind, hep-th/9611164; O.J. Ganor, S. Ramgoolam and W. Taylor,
hep-th/9611202.}
\REF\THO{G. 't Hooft, Nucl. Phys. {\bf B138} (1978) 1; Nucl. Phys. {\bf B153}
(1979) 141.}
\REF\GR{Z. Guralnik and S. Ramgoolam, hep-th/9702099.}
\REF\ROZ{M. Rozali, hep-th/9702136.}
\REF\IMA{Y. Imamura, hep-th/9703077.}
\REF\MST{J. Maldacena and A. Strominger, hep--th/9609026.}
\REF\MALD{J. Maldacena, hep-th/9607235.}
\REF\VAF{C. Vafa, Nucl. Phys. {\bf B463} (1996) 415, hep-th/9511026; Nucl.
Phys. {\bf B463} (1996) 435, hep-th/9512078.}
\REF\GOP{R. Gopakumar, hep-th/9704030.}
\REF\DV{R. Dijkgraaf, E. Verlinde and H. Verlinde, Nucl. Phys. {\bf B486}
(1997) 77, hep-th/9603126; Nucl. Phys. {\bf B486} (1997) 89, hep-th/9604055.}
\REF\PP{J. Polchinski and P. Pouliot, hep--th/9704029.}
\REF\FOUR{J. Maldacena and A. Strominger, Phys. Rev. Lett. {\bf 77} (1996) 428,
 hep-th/9603060; C. Johnson, R. Khuri and R. Myers, Phys. Lett. {\bf B378}
(1996) 78, hep-th/9603061.}
\REF\BSS{T. Banks, N. Seiberg and S. Shenker, hep-th/9612157.}
\REF\LE{L. Susskind, hep-th/9309145.}
\REF\HRS{E. Halyo, A. Rajaraman and L. Susskind, hep-th/9605112.}
\REF\HKRS{E. Halyo, B. Kol, A. Rajaraman and L. Susskind, hep-th/9609075.}
\REF\STR{L. Motl, hep-th/9701025; T. Banks and N. Seiberg, hep-th/9702187; R.
Dijkgraaf, E. Verlinde and H. Verlinde, hep-th/9703030.}
\REF\HP{G. Horowitz and J. Polchinski, hep-th/9612146.}
\REF\DPS{M. Douglas, J. Polchinski and A. Strominger, hep-th/9703031; J.
Maldacena, hep-th/9705053.}
\REF\HAL{E. Halyo, hep-th/9610068; hep--th/9611175.}


%
%

\chapter{Introduction}

During the last year and a half our understanding of the microscopic origin of
black hole entropy
increased enormously. This was achieved by considering $D=5$ ($D=4$) black
holes with
three (four)  RR charges and finite entropy at extremality[\SV,\CM]. The main
tool that enabled the entropy counting was D brane technology[\DB]. It was
shown that these black holes were built out of different
types of D branes (and other charges such as momentum)[\HMS]. In this picture,
the microscopic degrees of freedom are the momentum modes which are carried by
the massless open strings stretched between the different branes. It was later
realized that in the strongly interacting black hole regime
the system corresponds to a long string with fractional momentum modes[\MS].
The above picture can also be described in terms of the SYM theory which
describes the world--volume of the D branes.
In that description the black hole degrees of freedom correspond to the flat
directions of the SYM theory. Surprisingly, it was found that this picture can
be generalized to the near--extreme cases
when it is not protected by the supersymmetric nonrenormalization
theorems[\MAL]. Moreover,
Hawking radiation of different types of  scalars from black holes was shown to
be reproduced in this framework[\DM,\SCA].

Since string theory can be derived from M theory one would like to understand
black
holes in different dimensions from an M theoretical point of view. Already
(extreme and nonextreme)
black holes in different dimensions have been identified as configurations of
intersecting M branes[\KT]. For example, the five dimensional (extreme) black
hole
is described by five branes intersecting membranes over a string with momentum
flowing on it.
The nonextreme black hole is obtained by also allowing antimomentum to flow
along the intersecting string.
The only candidate for the nonperturbative formulation of M theory is the
M(atrix) theory[\BFSS]. In this framework,
M theory in the infinite momentum frame is described by infinitely many
longitudinal momentum
modes ($ {\tilde D0}$ branes) and their interactions due to open strings
stretched between them.
In order to obtain the eleven dimensional description one takes the limit $N
\to \infty, R \to \infty$
with $p=N/R \to \infty$. The Lagrangian is given by the dimensional reduction
of the ${\cal N}=1$ $U(N)$ SYM theory in $D=10$ to $0+1$ dimensions[\WIT].
Matrix thoery compactified on a torus $T^d$ is described by the
dimensional reduction of the above SYM theory to $d+1$ dimensions[\WATI,\BFSS].
However, there is another
formulation of the matrix model in the light--cone gauge[\LEN]. In this case,
the Lagrangian and the degrees of freedom are those in the infinite momentum
frame but $p,N$ and $R$ are finite. Finite $R$ is essential for our purposes
because we want to describe $D=5$ black holes by compactifying the matrix model
only  on $T^5$ (rather than on $T^6$ since not much is known about this case).
When $R \to \infty$ the five dimensional black hole becomes a black string in
six dimensions[\HS].
Finite $N$ is also relevant since for black hole configurations with
longitudinal momentum
$N$ is one of the charges and needs to be finite.

Black holes in the framework of matrix model were considered
recently[\DVV,\LM]. These were black holes with momentum flowing in the
light--cone direction.  It was shown that one can obtain the correct energy and
entropy for these black holes in terms of either an effective string [\LM] or
the world--volume of a NS five brane[\DVV]. This was done in the infinite
momentum frame where $N \to
\infty$. As a result, these works considered cases with infinite momentum but
finite momentum density such that $R \sim N^{1/2}$. However, there are a number
of problems with black holes that carry longitudinal momentum. First, one of
the charges $N$ diverges which is not the case for a finite black hole. Second,
nontrivial effects can take place in the limit $N \to \infty$ compared to the
better understood finite $N$ case. Third, in this case one does not know what
energy to assign to the rank of the $U(N)$ SYM gauge group (which corresponds
to the ${\tilde D0}$ branes) from the SYM point of view.
Finally, this is not the only configuration which realizes the $D=5$ black
hole.
There are other configurations which are related to the above by rotations in M
theory and by U dualities in IIA string theory.

In this paper, we consider $D=5$ black holes with three charges from the matrix
theoretical point of view. This is done in the finite $N$ formulation of the
matrix model which is in the light--cone gauge rather than the original
infinite momentum frame. The reason for this
is that for finite $N$ the light--cone direction is also compact giving the
sixth compact direction
required for a matrix  model description. For $N \to \infty$ our configurations
describe black strings in six dimensions.
We briefly mention configurations with momentum in the light--cone direction
since this case was
already examined. Our main interest is in matrix model configurations with
momentum in one of the transverse
directions. Matrix theory on $T^5$ is described by $5+1$ $U(N)$ SYM theory and
therefore
the black hole is a particular configuration of this theory. (For another
description of matrix theory on $T^5$ see [\BRS] where  it is argued that this
is a tensor theory with $(2,0)$ supersymmetry.
In any case, the SYM theory can be taken as the effective low energy theory in
the box even if
it is not a complete description including ultraviolet effects.)
We identify the BPS states of the SYM theory such as electric and magnetic
fluxes, instantons, momenta etc. [\FHRS]
and find configurations which correspond to black holes with transverse
momentum charge. These have four BPS charges, the fourth one being the rank of
the gauge group which gives the longitudinal momentum of the configuration. We
derive a formula for the entropy of these SYM configurations in the box which
is not completely U dual (as it does not contain the transverse five brane
charge).
We find that the rank does not enter the entropy of the black hole for the
configurations
with transverse momentum. This shows the invariance of black hole entropy under
boosts in
the light--cone direction as expected. In our opinion these configurations give
a clearer picture of five dimensional black holes than the ones with
longitudinal momentum. We also calculate the energy of these configurations in
the SYM theory which corresponds to the light--cone energy of the black hole.
The mass and entropy we find match those of the semiclassical black hole
precisely. We extend our calculations to nonextreme black holes in the dilute
gas approximation again finding agreement. One case that we cannot apply our
results is the configuration with transverse five branes since they do not have
a description only in terms of the SYM variables [\EDI,\BRS] and our entropy
formula does not contain their charge.

The outline of the paper is as follows. In section 2, we consider the matrix
model compactified on
$T^5$. This is described by a $5+1$ $U(N)$ SYM theory and we find all its BPS
states which
constitute the black hole. Section 3 is a review of five dimensional black
holes with three charges. We consider four different ways of realizing these
black holes in matrix theory.
In section 4, we find the SYM configurations which correspond to the above
realizations. We
calculate the mass and entropy of the black holes from the SYM theory and show
that
they match the semiclassical black hole results. We also consider nonextreme
black holes in the dilute
gas approximation. In section 5, we discuss our results and their possible
implications.

\chapter{M(atrix) Theory on $T^5$}

In this section, we review some facts about the matrix model compactified on a
five torus.
This will be essential for describing black holes in $D=5$ since the five
toroidally compactified
dimensions together with the light--cone direction (which is compact for finite
$N$) will give us the six compact dimensions of M theory. In eleven
(noncompact) dimensions the matrix model
is described by carriers of longitudinal momentum ($\tilde{D0}$ branes) with
the Lagrangian [\BFSS]
$$L = Tr \left({1 \over 2R} (D_0X^i D_0X^i)-{1 \over 4R} [X^i,X^j]^2+ fermionic
\quad terms\right)          \eqno(1)$$
where $R$ is the length of the light--cone direction, $X^i$ are $N \times N$
matrices, $D_0=\partial_0+iA_0$ and $i=1, \ldots,9$.

Matrix theory compactified on $T^d$ is described by a $d+1$ dimensional $U(N)$
SYM theory
on a $d$ dimensional finite box[\WATI,\FHRS]. This theory is obtained by
dimensionally reducing an ${\cal N}=1$
SYM theory in $9+1$ dimensions to $d+1$ dimensions.
For the $T^5$ compactification, we consider the $5+1$ dimensional SYM theory
with the Lagrangian
$$L=\int^V d^5\sigma \quad Tr \left(-{1\over 4g_6^2}F_{\mu \nu}^2+(D_{\mu}
X^i)^2+g_6^2[X^i,X^j]^2+fermionic \quad terms \right)    \eqno(2)$$
where $V=\Sigma_1 \Sigma_2 \Sigma_3 \Sigma_4 \Sigma_5$ is the volume of the
box,
$g_6$ is the gauge coupling constant, $\mu,\nu=0, \ldots, 5$ and
$i,j=6,\ldots,9$.

The SYM theory is in a five dimansional box with sides (parametrized by
$\sigma_1,\ldots, \sigma_5$) of length[\ST]
$$\Sigma_i={\ell_{11}^3 \over {RL_i}} \eqno(3)$$
(where the transverse compact dimensions are of length $L_1,L_2,L_3,L_4,L_5$
and the light--cone direction is of length $R$).
The dimensionful gauge coupling constant is[\EDI]
$$g_6^2={\ell_{11}^9 \over{R^2 L_1 L_2 L_3 L_4 L_5}} \eqno(4)$$

The $D=5$ IIA string theory has 27 point--like BPS states which are in the
fundamental representation of the U duality group $E_6$. We will identify these
BPS states in the
SYM theory.  These BPS states correspond to the constituents of the different
$D=5$ black holes with three charges.
The $5+1$ SYM theory has only 16 BPS states which are the five electric fluxes
(the Kaluza--Klein momenta), the ten magnetic fluxes (the wrapped
membranes)[\FHRS]
and a magnetic flux through a plane which is not manifest in the box (the
wrapped transverse five brane)[\EDI].
When one considers a compact longitudinal
direction (finite $N$) one can also have five instantonic strings (the
longitudinal five branes),
five momenta in the SYM box (the longitudinal membranes)
and the rank $N$ of the $U(N)$ SYM theory (the longitudinal momenta).
Altogether these
are 27 BPS states of the $D=5$ string theory.

The five Abelian electric fluxes which correspond to Kaluza--Klein modes
are given by[\GR]
$$\epsilon_{ijklm}E_i^A \Sigma_j \Sigma_k \Sigma_l \Sigma_m={2\pi n_i \over N}
1_{N \times N} \eqno(5)$$
This is accompanied by a non--Abelian electric flux
$$\epsilon_{ijklm}E_i^{NA} \Sigma_j \Sigma_k \Sigma_l \Sigma_m={2\pi n_i \over
N}  \omega
\eqno(6)$$
where $\omega=diag(1,1,1, \ldots,1-N)$ is an $SU(N)$ matrix. Similarly, the ten
Abelian
magnetic fluxes which describe the wrapped membranes are given by[\THO,\GR]
$$F_{ij}^A \Sigma_i \Sigma_j={2\pi n_{ij} \over N} 1_{N \times N} \eqno(7)$$
with the non--Abelian fluxes
$$F_{ij}^{NA} \Sigma_i \Sigma_j={2\pi n_{ij} \over N} \omega \eqno(8)$$
The wrapped transverse five brane is given by the Abelian magnetic flux[\EDI]
$$F_{5\sigma}^A \Sigma_5 \Sigma={2\pi n \over N} 1_{N \times N} \eqno(9)$$
and a non--Abelian flux obtained from the above by substituting $\omega$ for
the unit matrix.
Here $\Sigma=g_5^2=\ell_{11}^6/R L_1 L_2 L_3 L_4$ is the size of a new
direction
(parametrized by $\sigma$) which opens up as $g_5^2 \to \infty$
but is not manifest in the box[\ROZ].
The light--cone energy of these states are reproduced only by the Abelian
fluxes.
The form of $\omega$ changes when there are two orthogonal magnetic fluxes or
magnetic and elextric fluxes which have a common direction. In these cases, the
non--Abelian
fluxes also contribute to the energy as we will later see in the black hole
context. Note
that for each BPS state the amount of Abelian and non--Abelian fluxes are
equal.
The five longitudinal membranes are described by photons in the box with
momenta[\IMA]
$$p_i={2\pi m_i \over \Sigma_i} \eqno(10)$$
The five longitudinal five branes are given by the instantonic strings (say
along $\sigma_5$) with energy $n/g_5^2$
or tension $n/g_6^2=n/ \Sigma \Sigma_5$.
The last BPS state is given by the rank $N$ of the $U(N)$ gauge group and
corresponds to the
$\tilde{D0}$ branes of matrix theory with mass $n/R$.

\bigskip
\chapter{Five Dimensional Black Holes}

In this section, we review the solution for the $D=5$ black hole with three
charges[\HMS].
The classical solution to the low energy equations of motion in type
IIB string theory
compactified on $T^5$ is given by the metric $g_{\mu \nu}$, the RR
antisymmetric tensor
$B_{\mu \nu}$  and the dilaton $g^2=e^{-2 \phi}$.
The NS three form, the self--dual five form and the RR scalar are set to zero.
Also, the
asymptotic value of the dilaton $\phi$ is taken to be zero.
The classical five dimensional (nonextreme) black hole metric is given by
$$ds^2=-f^{-2/3} \left(1-{r_0^2 \over r^2} \right)dt^2+f^{1/3} \left[\left(1-
{r_0^2 \over r^2} \right)^{-1}dr^2+r^2d\Omega_3^2 \right] \eqno(11)$$
where
$$f=\left(1+{{r_0^2 sinh^2 \alpha}\over r^2} \right) \left(1+{{r_0^2 sinh^2
\beta} \over r^2} \right)
\left(1+{{r_0^2 sinh^2 \gamma} \over r^2} \right) \eqno(12)$$
The solution is parametrized by six parameters, $\alpha, \beta,\gamma, r_0$ and
the compactified one and four volumes $2 \pi R$ and $(2 \pi)^4 V$.
The total energy of the black hole is
$$E={RV r_0^2 \over {2g^2 \alpha^{\prime 4}
}}(cosh 2\alpha+ cosh 2\beta + cosh 2 \gamma) \eqno(13)$$
The entropy of the black hole is found from the area of the horizon using the
Bekenstein--Hawking formula
$$S={A_H \over {4G_5}}={2 \pi RV r_0^3 \over {g^2 \alpha^{\prime 4}}} cosh
\alpha~ cosh \beta~
cosh \gamma \eqno(14)
$$
where the ten and five dimensional Newton constants are given by
$
G_{10}=8 \pi^6 g^2 \alpha^{\prime 4}
$
and
$
G_5=G_{10} / (2\pi)^5 RV
$.
The black hole carries the three charges
$$\eqalignno{Q_5&={r_0^2\over{2 g\alpha^{\prime}}} sinh (2 \alpha)
&(15a)
 \cr
Q_1&={V r_0^2\over{2 g \alpha^{\prime 3}}} sinh (2 \beta) &(15b)\cr
n&={R^2V r_0^2\over{2 g^2 \alpha^{\prime 4}}} sinh (2 \gamma) &(15c)}$$
These are the charges of the black hole under the RR three form $H_3$, its
dual $H_7$ and Kaluza--Klein two form coming from the metric.
The extreme black hole limit is obtained by $r_0 \to 0$ and $\alpha, \beta,
\gamma \to \infty$
with the charges $Q_1,Q_5,n$ fixed.  For the nonextreme case we will consider
only the
dilute gas approximation which holds for $R>> \sqrt{\alpha^{\prime}}$, $V \sim
\alpha^{\prime 2}$
and all charges of the same magnitude, i.e. $Q_5 \sim Q_1 \sim n$[\MST]. This
is the region of the parameter space in which
nonextreme entropy can be reliably calculated. In this regime, when energy is
added to the black hole
very few anti--D five and one branes are excited compared to the number of
antimomenta. Therefore, the dominant contribution to the entropy change comes
from the momentum modes.

The properties of the black hole can be written in a suggestive way if we trade
the parameters $\alpha, \beta,\gamma, r_0,R,V$ for $N_5,N_1,n_L,n_R$ defined by
$$\eqalignno{ N_5&={ r_0^2 \over {4 \alpha^{\prime}}} e^{2 \alpha} &(16a)\cr
                                    N_1&={V r_0^2 \over {4g^2 \alpha^{\prime
3}}} e^{2 \beta} &(16b)\cr
                                     n_L&={R^2V r_0^2 \over {4g^2
\alpha^{\prime 4}}} e^{2\gamma}&(16c)\cr
                   n_R&={R^2V r_0^2 \over {4g^2 \alpha^{\prime 4}}} e^{-2
\gamma} &(16d)}$$
In terms of the above brane numbers numbers, the charges of the black hole are
$Q_1=N_1$,
 $Q_5=N_5$, $n=n_L-{n_R}$. The black hole mass is
$$M_{BH}={N_5 RV \over {g \alpha^{\prime 3}}}+
{ N_1 R\over {g \alpha^{\prime}}}  + {1 \over R}(n_L+n_R)
\eqno(17)$$
The entropy can be written as
$$S=2\pi \sqrt{N_5 N_1} (\sqrt{n_L}+
\sqrt{n_R}) \eqno(18)$$
The extreme limit corresponds to $n_R=0$.

The microscopic description of the black hole in terms of D branes is as
follows[\MALD]. An extreme
black hole with charges $Q_5$, $Q_1$ and $n$ is described at weak coupling by
$Q_1$
D one branes inside $Q_5$ D five branes with $n$ units of momentum along the D
string. The D strings are confined inside the world--volume of the D five
branes and therefore have oscillations in only the four transverse directions.
The system can be described by a
configuration with one long string of length $Q_5 Q_1 R$ which is preferred for
entropy reasons[\MS].
This can be interpreted as fractionation of momentum along the string and leads
to the correct
black hole entropy. From the point of view of the world--volume theory of the D
strings this is
a SYM theory in $1+1$ dimensions which is described by a CFT with central
charge
$c=6 Q_5 Q_1$ and total momentum $n$. The entropy of this system is given by
(for $c>>n$)[\SV]
$$S=2\pi \sqrt{cn \over 6} \eqno(19)$$
which reproduces the correct black hole entropy. For extreme black holes this
weak coupling counting can be reliably extrapolated to strong coupling which is
the black hole regime due to supersymmetry. Amazingly the same can be done for
nonextreme black holes in the dilute gas approximation and for low
energies[\MAL].

We saw that the $D=5$ black hole in IIB string theory is given by D one branes
inside D five branes with momentum along it. In order to make direct contact
with M theory we should T
dualize along one direction and pass to a configuration in IIA string theory.
If we T dualize along a direction inside the four volume $V$ we get four branes
intersecting membranes on a string
along which momentum flows. On the other hand, we can T dualize along the $R$
direction and
obtain zero branes inside four branes with wound strings orthogonal to the four
brane. There are
other IIA configurations which are related to these by U dualities.
The above configurations in IIA string theory are described in eleven
dimensional M theory by $N_5$ five branes intersecting $N_2$ membranes over a
string on which there is $N_0$ units of momentum[\KT]. Since there are six
compact dimensions
of the matrix model (the five torus and the light--cone) there are a number of
ways this configuration can be realized. We will consider four possibilities
which are related by rotations in M theory and by various U dualities in five
dimensional IIA string theory.
The four cases are as follows:

1) In this case the ($N_5$) five branes and the ($N_2$) membranes are
longitudinal with ($N_0$ units of) momenta in the light--cone direction. This
configuration and its S dual have been considered in refs. [\DVV] and [\LM]
respectively. In the notation of [\LM] it is given by
$$\left[
\matrix{
11& 10& 9& 8& 7& .\cr
11& .& .& .& .& 6 \cr
p& .& .& .& .& .\cr
}
\right] $$
Here the first, second and third rows describe the orientation of the five
branes, membranes and momenta respectively. 11 denotes the light--cone
direction and 10
is the direction related to the IIA string coupling constant. The momentum
modes are the $\tilde{D0}$ branes of matrix theory. In the $5+1$ SYM theory
this
configuration corresponds to a $U(N_0)$ theory with $N_5$ instantonic strings
and $N_2$
units of momenta in the $\sigma_6$ direction of the box.

2) This configuration is given by $N_5$ longitudinal five branes, $N_2$
membranes along
the 10 and 6 directions and $N_0$ units of momenta along the 10 direction.
Schematically
$$\left[
\matrix{
11& 10& 9& 8& 7& .\cr
.& 10& .& .& .& 6 \cr
.& p& .& .& .& .\cr
}
\right] $$
Now, momentum modes are the zero branes of IIA string theory. In the SYM theory
this
configuration is described by four charges; the three above and a fourth ($N$)
one which is the
number of  $\tilde{D0}$ branes or the longitudinal momentum of the whole
system. Thus, we
have a $U(N)$ gauge theory with $N_5$ instantonic strings along the $\sigma_6$
direction, $N_2$ units of magnetic flux in the $\sigma_6 \sigma_{10}$ plane
and $N_0$ units of electric flux in the $\sigma_{10}$ direction.

3) Another possibility is the configuration
$$\left[
\matrix{
11& 10& 9& 8& 7& .\cr
.& .& 9& .& .& 6 \cr
.& .& p& .& .& .\cr
}
\right] $$
This is similar to the second case but it is a different configuration of IIA
string theory. Case 2 describes four branes, membranes and strings which was
obtained by T dualizing the black hole
along the $V$ direction. Case 3, on the other hand, corresponds to the case
obtained by T dualizing along the $R$ direction. The SYM
picture is obtained by $9 \leftrightarrow 10$ from the above description.

4) Finally, one can also have  transverse five branes, longitudinal membranes
and momenta
along the intersecting string which is along a transverse direction
$$\left[
\matrix{
.& 10& 9& 8& 7& 6\cr
11& .& 9& .& .& . \cr
.& .& p& .& .& .\cr
}
\right] $$
This case is problematic due to the fact that the transverse five brane cannot
be expressed solely in terms of the SYM variables in the five dimensional box.

Note that the first configuration has only three charges whereas the others
have four charges. The fourth charge corresponds to the rank of the gauge group
(the number of $\tilde{D0}$ branes or light--cone momentum). In the next
section, we will see that the black hole entropy for cases 2 and 3
is independent of the rank of the group. This is a satisfactory result because
it shows that entropy is
independent of longitudinal momentum as it should be. Of course, for the first
case entropy
depends on the rank of the group since there are only three charges and one of
them is the momentum charge of the black hole.
In our opinion, cases 2 and 3 are better descriptions of the black hole than
case 1 due to the manifest invariance of
the black hole entropy under longitudinal boosts. Also in case 1 one does not
know what energy to assign to the rank of the gauge theory in a box.
In cases 2 and 3 there are three BPS states in the SYM theory and they
correctly reproduce
the mass of the black hole. Another problem with case 1 is that antimomentum
states are
not allowed since we are in the IMF or the light--cone gauge. This would
correspond to a negative rank for the gauge group which has no meaning. This is
not a problem in cases 2 and 3 since all three charges in the SYM theory can be
negative.

\bigskip
\chapter{M(atrix) Black Holes in Five Dimensions}

In this section, we calculate the entropy and energy of the SYM configurations
we considered in the
previous section which correpond to the five dimensional black hole. The first
case has
been considered in refs. [\DVV] and [\LM] and we will discuss it very briefly.
Our main interest will be
in cases 2 and 3.  In order to calculate the entropy of the SYM configurations
with three
different BPS charges, we will first consider the $4+1$ SYM theory and then
translate our
formula for the entropy to the $5+1$ SYM theory. The energy of the SYM
configuration calculated from eq. (2) corresponds to the light--cone energy of
the black hole.

The entropy of the $4+1$ dimensional $U(N)$ SYM theory with instanton number
$k$ and momentum $p$ is well known. This system is described by a
superconformal sigma model with the target space $S^{Nk} T^4$ and thus has a
central charge $c=6 Nk$ (since there are four bosonic and four fermionic
oscillations)[\VAF,\SV]. The entropy of the system is given by the usual
formula
$$S=2\pi \sqrt{c p \over 6}= 2\pi \sqrt{Nkp} \eqno(20)$$
The BPS charges that enter this formula are the rank of the gauge group (four
branes),
instanton number (zero branes) and momentum in the box.
One can generalize this formula to include the other BPS charges including the
longitudinal branes (but not the transverse five brane). In ref. [\THO] it was
shown that
when there are two orthogonal magnetic fluxes (membranes) one gets a fractional
instanton
number given by
$$\nu=k-{n_{ij} n_{kl} \over N} \eqno(21)$$
where $n_{ij}$ correspond to the units of magnetic flux in the $ij$ direction
as in eq. (7). The configuration in the
SYM theory becomes
$$F_{ij}^A \Sigma_i \Sigma_j={2\pi n_{ij} \over N} 1_{N \times N}  \qquad
F_{kl}^A \Sigma_k \Sigma_l={2\pi n_{kl} \over N} 1_{N \times N}
\eqno(22a,b)$$
with the corresponding non--Abelian fluxes
$$F_{ij}^{NA} \Sigma_i \Sigma_j={2\pi n_{ij} \over N} \omega \qquad
F_{kl}^{NA} \Sigma_k \Sigma_l={2\pi n_{kl} \over N} \omega
\eqno(23a,b)$$
where $\omega=diag(k, \ldots,k,-l,\ldots,-l)$ and $k+l=N$. These configurations
are exactly
't Hooft's toron configurations which describe fractional instantons.
In this case, both the Abelian and non--Abelian fluxes contribute to the energy
of the configuration. This system is known to be a superconformal sigma
model with target space $S^{N \nu} T^4$.

In addition, it was shown in ref. [\GOP] that when there are magnetic and
electric fluxes
with a common direction the momentum in the box becomes fractional
$$p = m_i-{n_j n_{ij} \over N} \eqno(24)$$
The SYM configuration is now given by the magnetic fluxes
$$F_{ij}^A \Sigma_i \Sigma_j={2\pi n_{ij} \over N} 1_{N \times N}  \qquad
F_{ij}^{NA} \Sigma_i \Sigma_j={2\pi n_{ij} \over N} \omega \eqno(25a,b)$$
and the electric fluxes
$$\epsilon_{ijklm}E_j^A \Sigma_i \Sigma_k \Sigma_l \Sigma_m={2\pi n_i \over N}
1_{N \times N} \eqno(26)$$
and
$$\epsilon_{ijklm}E_j^{NA} \Sigma_i \Sigma_k \Sigma_l \Sigma_m={2\pi n_i \over
N}  \omega
\eqno(27)$$
with the same $\omega$ as above. Once again the total energy is a sum over the
Abelian
and non--Abelian contributions.

When both instanton number and box momentum are fractional the system is
described by a
CFT with the target space $S^{N \nu} T^4$ and total momentum $p$. The obvious
generalization of the entropy formula in eq. (20) is
$$S=2\pi \sqrt{N|(k-n_{ij}n_{kl}/N)||(m_i-n_jn_{ij}/N)|} \eqno(28)$$

The BPS charges that appear in this formula are the rank $N$, magnetic fluxes
$n_{ij}$,
electric fluxes $n_i$, box momentum $m_i$ and instanton number $k$. Considering
the $4+1$ SYM
theory as the world--volume theory of a four brane these correspond to the
number of four branes and membranes, strings, longitudinal  membranes and zero
branes inside the four branes respectively. Eq. (28)
for the entropy can be easily generalized to the $5+1$ SYM case giving
$$S=2\pi \sqrt{N|(k_i-\epsilon_{ijklm}n_{jk}n_{lm}/N)||(m_i-n_jn_{ij}/N)|}
\eqno(29)$$
Considering the $5+1$ dimensional SYM  as the world-volume
theory of a D five brane the BPS charges $N, n_{ij}, n_i, m_i, k_i$ correspond
to five branes and three branes, strings, longitudinal
three branes and instantonic D strings inside five branes respectively. From
the point of view of the original brane configuration in M theory these are
${\tilde{D0}}$ branes, membranes, transverse momentum, longitudinal  membranes
and longitudinal five branes respectively. There is a special direction in the
SYM theory which is given by the direction of the instantonic string.
The two three branes inside the five brane and the three branes and strings
also intersect over the same direction $i$. This is the origin of the effective
string picture of the black hole in matrix theory. We see that using the SYM
theory in the box, we obtained the fractionation of momentum and instanton
number which is crucial for understanding black hole entropy. Since this is a
pure SYM result it can be seen as the matrix theoretical origin of
fractionation considering the connection between the compactified matrix theory
and SYM theories.

Note that all but one of the 27 BPS charges of the SYM theory we found in
section 2 appear in the above formula. The only exception is the transverse
five brane charge
which does not have a description only in terms of the SYM variables in the
box[\EDI,\BRS]. As a result, the entropy formula is not completely U dual, i.e.
$E_6$ symmetric. This also the reason why we cannot obtain the entropy of the
fourth configuration of the last section which includes  transverse five
branes. In ref. [\DVV] five dimensional black holes with NS charges were
examined
and a very similar formula was obtained for the entropy.
This case corresponds to the description of matrix theory on $T^5$ by the
$(2,0)$ tensor theory. In that case, the entropy of the configuration is due to
the noncritical strings which live on $T^5$
a theory which describes the world--volume theory of NS five branes.
These two descriptions are expected to give the same entropy since they are S
duals of each other.

We can now calculate the energy and entropy for the different descriptions of
the $D=5$ matrix black hole in terms of the $5+1$ SYM theory. Consider the
first case of the previous section.
We have  longitudinal five branes, longitudinal membranes and momenta along the
light--cone direction. In the $5+1$ SYM theory these are described by
instantonic strings ($N_5$),
box momenta ($N_2$) and the rank of the gauge group ($N_0$). This case was
studied in ref. [] and
will not be reviewed here.
The black hole entropy is found from eq. (29) (using $N=N_0, k=N_5, m_i=N_2$)
$$S_{BH}=2\pi \sqrt{N_0 N_5 N_2} \eqno(30)$$
This matches the semiclassical result given by eq. (18).

The second case is more interesting for our purposes. This configuration is
given by
$N_5$ longitudinal five branes intersecting $N_2$ transverse membranes and
$N_0$ units of momentum along
the 10 (the IIA string coupling constant) direction. In the SYM theory these
become instantonic strings, magnetic flux and
electric flux respectively. The light--cone momentum (number of $\tilde{D0}$
branes) is
$N$ and gives the rank of the gauge group $U(N)$. Now, we have three BPS states
in the SYM theory and we can find their energies. The $N_5$ instantons have
energy
$$H_{inst}={1\over g_6^2} Tr \int^V d^5 \sigma ~ F_{\mu \nu} \tilde{F}_{\mu
\nu}={N_5 \over g_5^2} \eqno(31)$$
The energy of the electric and magnetic fluxes is given by the sum of the
Abelian and non--Abelian parts, $H^A+H^{NA}$ where
$$\eqalignno{H^A&={1 \over 2g_6^2} Tr \int^V d^5 \sigma ~ (F^A_{6,10})^2
+{g_6^2 \over 2} Tr \int^V d^5 \sigma ~ (E^A_{10})^2 &(32a) \cr
&={2\pi^2 \over N g_6^2} N_2^2 { {\Sigma_7 \Sigma_8 \Sigma_9} \over {\Sigma_6
\Sigma_{10}}}
+{2\pi^2 g_6^2 \over N} N_0^2 {\Sigma_{10} \over  {\Sigma_6 \Sigma_7 \Sigma_8
\Sigma_9} }
&(32b)} $$
and
$$\eqalignno{H^{NA}&={1 \over g_6^2} Tr \int^V d^5 \sigma ~ F_{0,10}^{NA}
F_{6,10}^{NA}&(33a) \cr
&={4\pi^2 N_2N_0 \over N} {1 \over \Sigma_6} &(33b)}$$
Here we used the expressions for the electric and magnetic fluxes given by eqs.
 (5-8)
with $n_{6,10}=N_2$, $n_{10}=N_0$.
The total energy of the configuration is
$$\eqalignno{H&=H_{inst}+H^A+H^{NA} &(34a) \cr
&={N_5 \over g_5^2}+ \left({2\pi^2 \over N g_6^2}\right) {\Sigma_{10} \over
\Sigma_6 \Sigma_7 \Sigma_8 \Sigma_9}
\left({\Sigma_7 \Sigma_8 \Sigma_9 \over \Sigma_{10}} +g_6^2 \right)^2 &(34b)}$$
This should be equal to the light--cone energy of the black hole.
Using  the definitions of $\Sigma_i$, $g_6^2$  and the relations $\ell_{str}=
\ell_{11}^3/L_{10}$ and
$g_{str}^2=L_{10}^3/\ell_{11}^3$ it is easy to show that the total SYM energy
of the configuration is precisely the light--cone energy of the black hole.
Note that the mass of the five branes
is not squared since they are longitudinal whereas the membrane and zero brane
masses are squared since they are transverse.

The entropy of the SYM system is given by eq. (29)
$$S=2\pi \sqrt{N N_5 (N_0 N_2/ N)} \eqno(35)$$
which is precisely the entropy of the black hole. Note that the entropy does
not depend on the light--cone momentum or $N$ as it should be.
The third case can be obtained from the above by the interchange $9
\leftrightarrow 10$
and gives the same results.

We can now consider nonextreme black holes in the dilute gas approximation. In
order to do so we need to add a small amount of antimomentum (right--handed in
the notation of section 3) to
the black hole configuration. In terms of the SYM configuration this
corresponds to adding
a small amount of electric flux in the negative 10 direction such that ${\bar
E_{10}} << E_{10}$
for both the Abelian and the non--Abelian fields. The total electric field is
now $E_{tot}=E_{10}-
{\bar E_{10}}$. As a result of the negative electric flux  ${\bar E_{10}}$
there is both left and right--handed momentum in the SYM configuration
$$p_L={n_i n_{ij} \over N} \qquad p_R={{\bar n_i}n_{ij} \over N} \eqno(36)$$
where $n_i,{\bar n_i}$ correspond to $E_{10},{\bar E_{10}}$ in the notation of
eq. (5) respectively.
The system is now described by a CFT with the target space $S^{N \nu} T^4$ with
fractional $p_L$ and $p_R$. The entropy is
$$S=2\pi \left(\sqrt{c_L p_L \over 6}+\sqrt{c_R p_R \over 6} \right)
\eqno(37)$$
with $c_L=c_R=6$.

\bigskip
\chapter{Discussion and Conclusions}

In this paper, we examined configurations of the matrix model which correspond
to five dimensional black holes.  This was done in the light--cone gauge
formulation of the model with finite longitudinal momentum and radius of the
eleventh dimension.
We considered three different configurations that correspond to $D=5$ black
holes: one with  longitudinal and two with transverse momentum.
We obtained a formula for the entropy of  BPS configurations (with three
charges)
of the $5+1$ dimensional SYM theory which describes the matrix model
compactified on $T^5$.
The formula is not completely U dual since it does not contain the transverse
five brane charge.
Fractionation of momentum and/or instanton number which plays a crucial role in
understanding black hole entropy arises automatically from the SYM picture.
This can be seen as the origin of fractionation in the  (compactified) matrix
model.
We also calculated the mass of these configurations and found that the
energy and entropy of these match those of the black holes precisely.
We found that the entropy for the transverse momentum case does not depend on
the longitudinal momentum or $N$. When one takes the limit $N \to \infty$
(since entropy etc. do not depend on $N$) the black hole becomes a black string
in six dimensions. We generalized
our results for the nonextreme black holes in the dilute gas approximation.
Once again the mass
and entropy of the SYM configurations match the black hole results.

The black hole configurations we studied seem to manifest part of the $D=11$
Lorentz invariance of the matrix model[\PP].
First, as we saw longitudinal boost invariance was manifest for the black hole
configurations
with transverse momentum. The difficult part of the eleven dimensional Lorentz
invariance is the rotational symmetry between the longitudinal and transverse
dimensions. This requires that a given configuration boosted along the
light--cone direction is equivalent to the same configuration rotated to
a transverse direction (e.g. $9 \leftrightarrow 11$) and boosted along it. This
seems to be realized by our black hole configurations as can be seen by
comparing case 1 with cases 2 and 3. The mass, charge and entropy of these
configurations are the same and therefore this is a manifestation of eleven
dimensional Lorentz invariance.

As noted above, the entropy formula we derived for the SYM configurations which
correspond to the black holes
is not U dual or $E_6$ symmetric. In particular the transverse five brane
charge is missing from the formula. The present description of the five brane
requires the use of a new dimension which is not geometrically manifest. Thus,
the five brane cannot be described by the SYM and the box variables alone. In
order to obtain the entropy of case 4, what seems to be required is either a U
dual formula together with a SYM description of the five brane or a
generalization of our entropy formula to include the nonmanifest direction.
Note that the other description of the transverse five brane in the tensor
theory also involves the nonmanifest direction $\sigma$[\BRS].

Extending our results to four dimensional black holes does not seem
staightforward[\FOUR]. First, $D=4$
black holes require six branes (with finite mass) which do not have a
description in the matrix model. (Six branes which are built out of three
orthogonal stacks of membranes are known but these do not have finite
energy[\BSS].)
Second, four dimensional black holes require compactifying matrix model on a
six torus $T^6$ which is problematic.
In ref. [\FHRS] this issue was investigated and it was found that there is no
weakly coupled
desciption for the $T^6$ compactifications of the matrix model. This may be
related to the lack of
a superconformal fixed point in $6+1$ dimensional field theory.

We considered the nonextreme black holes only in the dilute gas approximation
(just as in the
D brane picture). Can we go beyond this approximation and consider
Reissner--Nordstrom black holes
for which the deviation from extremality is for the three charges
simultaneously? In previous work on matrix black holes and case 1 above this is
not possible since one of the charges is
the longitudinal momentum. The nonextreme case requires negative momentum which
corresponds to negative longitudinal momentum (anti ${\tilde D0}$ branes) which
is clearly not
possible. In the light--cone description there are no negative longitudinal
momentum modes and in the infinite momentum frame these modes decouple from the
system.
On the other hand, for configurations with momentum in the transverse direction
all three charges can be negative and there is no a priori reason why one
cannot realize the nonextreme Reissner--Nordstrom case.
This hopefully may be a way to go beyond the dilute gas approximation in a
reliable manner.

Taking the highly nonextreme (Schwarzschild) limit we may ask whether the
matrix model description of black holes enhances our understanding of black
hole entropy. In this limit, the D brane picture is not
helpful because Schwarzschild black holes are described by highly excited
fundamental strings
[\LE,\HRS,\HKRS].
For large number of antibranes, the branes and antibranes annihilate into
strings with no
charge. In M theory fundamental strings are wrapped membranes and therefore
states of the
SYM theory. The excited (oscillating) string is not a BPS state however. It is
interesting that a highly excited string becomes a black hole when the string
coupling is weak ($g<<1$ but
$gN^p >1$ for $p<1$) which is a well-understood limit of the compactified
matrix model[\STR].
The string oscillations
are described by the nonzero kinetic terms for the scalars in the SYM theory.
It seems that the highly excited string configuration can be examined in the
matrix model context.
On the other hand, the relation between mass and entropy of Schwarschild black
holes is known in any dimension. Therefore it would be interesting to consider
this limit in matrix model.

A related question is the transition between the weakly coupled collection of
SYM BPS states and the
strongly coupled configurations which describe the black hole.  In the string
and D brane context
this question was answered by the correspondence principle which states that
the transition to the black hole state occurs when the curvature in the string
metric is around the string scale[\HP].
Is there such a  principle in the matrix model in terms of the SYM variables?
This would require
the analogs of concepts such as the string length, black hole radius and
curvature etc. in the SYM picture. Unfortunately, the connection between the
space--time description of the black hole and the
SYM picture is not very clear. For example, among other things, the existence
and position of the horizon is not well--understood[\DPS].

One possible interpretation of the entropy formula we derived is as follows.
The black hole can be described as an effective (long) fundamental string with
a given momentum and winding charge and a rescaled tension (due to the presence
of a background five brane)[\HAL].
The momentum and winding are
fractionized and given eqs. (20) and (23). The factor $N$ arises due to the
rescaling of the string tension i.e. $\alpha^{\prime}_{eff} \to N
\alpha^{\prime}$. The effective fundamental string is dual to the real D
strings (inside the five branes) which are described by the instantons in the
SYM theory.

\bigskip
\centerline{Acknowledgements}
We would like to thank Lenny Susskind for very useful discussions.

\vfill

\refout

\end